\documentclass[aps, prl, reprint, superscriptaddress, showpacs,preprintnumbers,amsmath,amssymb,twocolumn]{revtex4-2}
\usepackage{amssymb}
\usepackage{amsmath}
\usepackage{bm}
\usepackage[dvips]{graphicx}
\usepackage{dcolumn}
\usepackage{longtable}
\usepackage{xcolor}
\usepackage[version = 4]{mhchem}

\newcommand{\casb}{CaSb$_2$}
\newcommand{\Tc}{$T_{\mathrm{c}}$}

\begin{document}
\title{Pressure evolution of the normal- and superconducting-state properties of the line-nodal material \casb ~revealed by $^{123}$Sb nuclear quadrupole resonance}
\author{H.~Takahashi}
\email{takahashi.hidemitsu.23r@st.kyoto-u.ac.jp}
\affiliation{Department of Physics, Graduate School of Science, 
Kyoto University, Kyoto 606-8502, Japan}

\author{S.~Kitagawa}
\affiliation{Department of Physics, Graduate School of Science, 
Kyoto University, Kyoto 606-8502, Japan}

\author{K.~Ishida}
\affiliation{Department of Physics, Graduate School of Science, 
Kyoto University, Kyoto 606-8502, Japan}

\author{A.~Ikeda}
\affiliation{Maryland Quantum Materials Center and Department of Physics,
University of Maryland, College Park, Maryland 20742, USA}
\affiliation{Toyota Riken-Kyoto University Research Center (TRiKUC), Kyoto 606-8501, Japan}

\author{S.~R.~Saha}
\affiliation{Maryland Quantum Materials Center and Department of Physics,
University of Maryland, College Park, Maryland 20742, USA}

\author{S.~Yonezawa}
\affiliation{Department of Physics, Graduate School of Science, 
Kyoto University, Kyoto 606-8502, Japan}

\author{J.~Paglione}
\affiliation{Maryland Quantum Materials Center and Department of Physics,
University of Maryland, College Park, Maryland 20742, USA}
\affiliation{Canadian Institute for Advanced Research, Toronto, Ontario, Canada M5G 1Z8}

\author{Y.~Maeno}
\affiliation{Department of Physics, Graduate School of Science, 
Kyoto University, Kyoto 606-8502, Japan}
\affiliation{Toyota Riken-Kyoto University Research Center (TRiKUC), Kyoto 606-8501, Japan}

\date{\today}

\begin{abstract}
\casb ~is the Dirac line-nodal material that exhibits a superconducting (SC) transition at 1.7 K.
In spite of its conventional SC state at ambient pressure, the transition temperature \Tc ~shows a peak structure against hydrostatic pressure.
We performed ac magnetic susceptibility and $^{123}$Sb nuclear quadrupole resonance (NQR) measurements on single-crystalline \casb ~under pressures up to 2.08 GPa.
\Tc ~monotonically increased in this pressure region, which is consistent with a previous study.
We observed continuous broadening of the NQR spectrum against pressure, which is a sign of unique compression behavior of the lattice.
In the normal state, the nuclear spin-lattice relaxation rate $1/T_1$ is proportional to temperature in all pressure values; typical of a metal.
However, $1/T_1T$ in the normal state is independent of pressure, indicating that the density of states at the Fermi energy $N(E_{\mathrm{F}})$, which is one of the parameters governing \Tc, is insensitive to pressure.
From these results, we conclude that $N(E_{\mathrm{F}})$ does not govern the origin of the enhancement in \Tc.
This is unusual for a weak electron-phonon coupling superconductor.
In the SC state, we revealed that the SC gap becomes larger and more isotropic under pressure.
\end{abstract}
\maketitle

\textit{Introduction.}
Materials with nontrivial topological features have been the central part of the research target in condensed matter physics. 
In particular, topological superconductivity has been attracting much attention since it could be a host of
exotic bound states. One of the promising candidates for bulk topological superconductivity is superconductivity in semimetals with nontrivial topological numbers.
In such materials, chemical doping is not necessary for the emergence of superconductivity. This ensures ``clean'' topological superconductivity. 
So far, many studies have been performed to discover superconductivity in topological semimetals such as Dirac \cite{hePressureinducedSuperconductivityThreedimensional2016,
PhysRevB.96.220506,PhysRevB.102.205117,https://doi.org/10.1002/adma.201901942,zhouBulkSuperconductivityDirac2021}, Weyl \cite{qiSuperconductivityWeylSemimetal2016,panPressuredrivenDomeshapedSuperconductivity2015,
liConcurrenceSuperconductivityStructure2017,Baenitz_2019,PhysRevB.99.020503,muPressureinducedSuperconductivityModification2021,PEI2021100509,PhysRevMaterials.5.084201,
PhysRevB.105.174502}, and line-nodal semimetals \cite{PhysRevB.93.020503,PhysRevB.100.064516,adamSuperconductingPropertiesTopological2021,shenTwodimensionalSuperconductivityMagnetotransport2020,
PhysRevResearch.4.L032004,PhysRevB.96.165123,PhysRevMaterials.5.074201,yamadaSuperconductivityTopologicalNodalline2021,ikedaSuperconductivityNonsymmorphicLinenodal2020}. 
Dirac and Weyl semimetals have topologically protected point nodes in their bulk electronic bands, while nodes form lines in the $\bm{k}$ space in line-nodal semimetals.
In terms of superconductivity, line-nodal semimetals seem to be especially advantageous due to their moderate density of states (DOS).
As a matter of fact, a greater number of superconducting (SC) line-nodal semimetals exhibit superconductivity without pressure or other techniques \cite{PhysRevB.93.020503,PhysRevB.100.064516,adamSuperconductingPropertiesTopological2021,
shenTwodimensionalSuperconductivityMagnetotransport2020,PhysRevResearch.4.L032004,PhysRevMaterials.5.074201,yamadaSuperconductivityTopologicalNodalline2021,ikedaSuperconductivityNonsymmorphicLinenodal2020}.
Therefore, the line-nodal semimetal is a good platform for exploring topological superconductivity and for understanding the relationship between nodal lines and superconductivity
in detail.

\begin{figure}[tbp]
    \includegraphics[width=8.65cm]{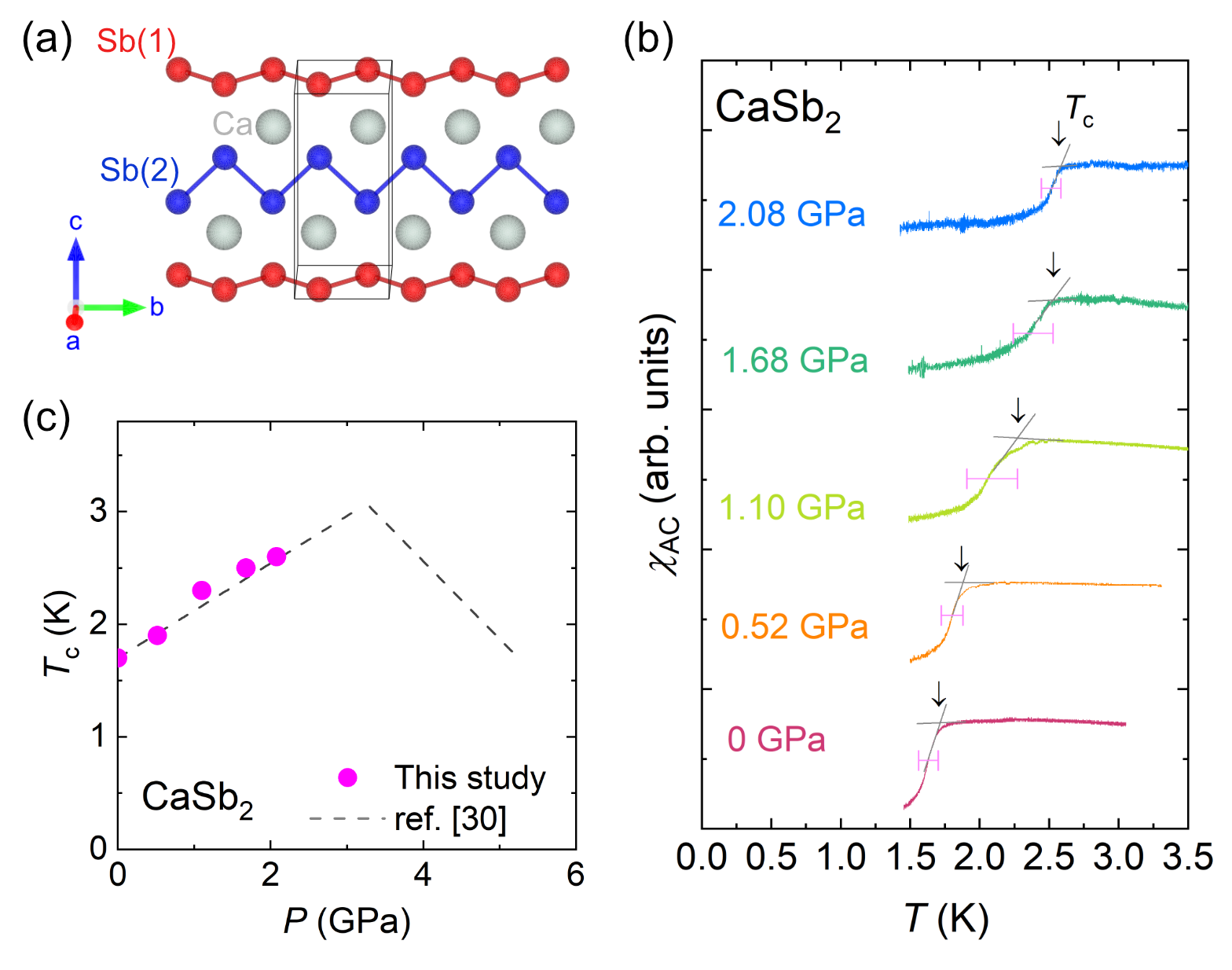}
    \caption{(Color online) 
    (a) Crystal structure of \casb ~with two nonequivalent Sb sites drawn by VESTA \cite{Momma:db5098}. (b) Temperature dependence of the ac susceptibility $\chi_{\mathrm{AC}}$ measured by a self-inductance method using a frequency range of 83.5-89.7 MHz under various pressure values. 
    (c) Pressure dependence of \Tc.
 The dashed lines are visual guidance identical to those in Ref. \cite{kitagawaPeakSuperconductingTransition2021}.}
    \label{fig:tc}
\end{figure}

\casb ~is a superconductor \cite{ikedaSuperconductivityNonsymmorphicLinenodal2020} which was predicted as a line-nodal material \cite{funadaSpinOrbitCouplingInducedTypeI2019}.
It crystalizes in a monoclinic structure with a nonsymmorphic space group ($P2_1/m$, No.~11, $C^2_{2h}$) with two nonequivalent crystallographic Sb sites, Sb(1) and Sb(2) [Fig.~\ref{fig:tc}(a)].
In most line-nodal materials, line nodes are gapped by spin-orbit coupling (SOC). However, the line nodes in \casb ~are protected against SOC by nonsymmorphic crystalline symmetry \cite{funadaSpinOrbitCouplingInducedTypeI2019},
making \casb ~suitable for investigating the superconductivity in line-nodal materials.

First-principles calculation predicts that \casb ~has two kinds of Fermi surfaces: a topologically trivial three-dimensional hole pocket that stems from Sb(2) and a pair of quasi-two-dimensional electron pockets 
related to the nodal lines that stems from Sb(1) \cite{takahashiSWaveSuperconductivityDirac2021}. Some of the authors of this Letter observed both of them by quantum oscillation measurements \cite{ikedaQuasitwodimensionalFermiSurface2022}.
Regarding superconductivity, we succeeded in observing the nuclear quadrupole resonance (NQR) signals of $^{121/123}$Sb(1) and 
revealed a full-gap $s$-wave SC behavior at ambient pressure \cite{takahashiSWaveSuperconductivityDirac2021}.
A full-gap behavior was also reported from magnetic penetration depth measurements \cite{PhysRevB.106.214521}.
Microfocused angle-resolved photoemission spectroscopy ($\mu$-ARPES) reported that the existence of electron pockets is sample dependent and pointed out that
topologically trivial hole pockets play a dominant role in superconductivity \cite{chuangFermiologyTopologicalLinenodal2022}.
These results indicate that, at least at ambient pressure, the SC state of \casb ~is conventional.

On the contrary, the deviation from the Bardeen-Cooper-Schrieffer (BCS) behavior in specific heat was reported \cite{PhysRevB.105.184504}.
Furthermore, we observed nonmonotonic behavior of the SC transition temperature $T_{\mathrm{c}}$ against hydrostatic pressure, 
indicating the unconventional character of the SC state in \casb ~\cite{kitagawaPeakSuperconductingTransition2021}.
The origin of the pressure dependence of \Tc ~remains unexplained.
We note that this kind of behavior in \Tc ~is also observed in a SC line-nodal metal NaAlSi \cite{schoopEffectPressureSuperconductivity2012}, whose origin is not revealed either.
We consider that it is important to clarify the origin of nonmonotonic change in \Tc ~to understand the SC properties of \casb ~and line-nodal materials, particularly those relating to band 
topology.

In this Letter, we report ac magnetic susceptibility measurements and NQR measurements on a $^{123}$Sb(1) site, at which the nodal behavior is expected, under hydrostatic pressures up to 2.08 GPa to reveal the origin of the increase in \Tc.
The increase in \Tc ~by pressure reported for polycrystalline samples was reproduced in single crystals.
In the normal state, the nuclear spin-lattice relaxation rate $1/T_1$ is proportional to temperature, namely the value of $1/T_1T$ is constant in all pressure values.
The constant value of $1/T_1T$, associated with DOS at the Fermi energy $E_{\mathrm{F}}$, is almost pressure independent.
This indicates that $N(E_{\mathrm{F}})$ (DOS at $E_{\mathrm{F}}$) is pressure independent.
From this result, we conclude that the change in \Tc ~is not ascribed to any changes in $N(E_{\mathrm{F}})$.
In the SC state, a higher coherence peak was observed under pressure, suggesting that anisotropy in the SC full gap is reduced under pressure.

\textit{Methods.}
We used single crystals of \casb ~with a slablike shape grown by the self-flux method as reported in Ref. \cite{ikedaQuasitwodimensionalFermiSurface2022}.
The ac magnetic susceptibility $\chi_{\mathrm{AC}}$ was measured by the NQR tank circuit to determine \Tc.
A standard spin-echo technique was used for the $^{123}$Sb NQR measurements (nuclear spin $I=7/2$).
The value of $1/T_1$ was obtained by measuring the time dependence of the spin-echo intensity after saturation of the nuclear magnetization $M(t)$.
The obtained data were fitted by the theoretical relaxation function \cite{PhysRevB.4.60}
\begin{align}
    m(t)=&\frac{2}{21}\exp\left(-\frac{3t}{T_1}\right)+\frac{25}{154}\exp\left(-\frac{10t}{T_1}\right)\notag\\
    &+\frac{49}{66}\exp\left(-\frac{21t}{T_1}\right)
\end{align}
in the same way as Ref.~\cite{takahashiSWaveSuperconductivityDirac2021} where $m(t)=[M(\infty)-M(t)]/M(\infty)$.
Pressure was applied using a NiCrAl-BeCu piston-cylinder-type cell with Daphne 7373 oil as the pressure medium.
The diameter of the initial sample space is 3.7 mm.
Pressure was determined by a Pb manometer using the relationship \cite{eilingPressureTemperatureDependence1981}
\begin{equation}
    P\ \mathrm{(GPa)}=\frac{7.18\ \mathrm{(K)}-T_{\mathrm{c}}\ \mathrm{(K)}}{0.364\ \mathrm{(K/GPa)}}.
\end{equation}

\textit{Results and Discussion.}
Figure \ref{fig:tc}(b) shows the temperature dependence of $\chi_{\mathrm{AC}}$ measured at various pressures.
A clear diamagnetic signal due to the SC transition was observed. The width of the transition, which is shown with an errorbar, becomes broader up to 1.10 GPa. 
The broadening of the SC transition can be attributed to the inhomogeneity of pressure, while the origin of the sharpening beyond 1.10 GPa is not clear.
\Tc ~is defined as the onset temperature of the SC transition.
The pressure dependence of $T_{\mathrm{c}}$ is plotted in Fig.~\ref{fig:tc}(c).
In this pressure range, $T_{\mathrm{c}}$ monotonically increases. This is consistent with the previous result 
shown with the dashed line \cite{kitagawaPeakSuperconductingTransition2021}.

\begin{figure}[tbp]
    \includegraphics[width=8.6cm]{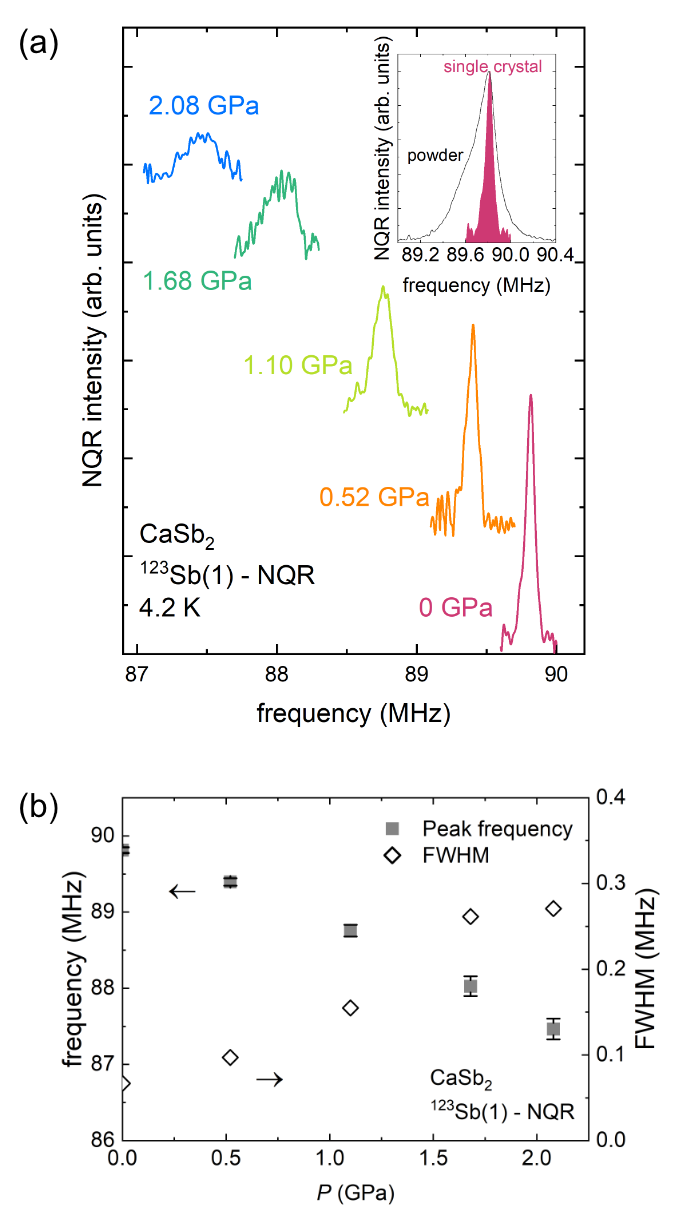}
    \caption{(Color online) 
    (a) $^{123}$Sb(1) NQR spectrum of \casb ~measured at 4.2 K under various pressure values. Inset: $^{123}$Sb(1) NQR spectrum at 0 GPa with that of the powder samples.
    (b) Pressure dependence of peak frequency and FWHM of the NQR spectrum.}
    \label{fig:spectra}
\end{figure}

We observed a $^{123}$Sb(1) NQR spectrum corresponding to the $\pm 3/2\leftrightarrow\pm 5/2$ transition.
The site assignment of the Sb(1) NQR signal was described in a previous study \cite{takahashiSWaveSuperconductivityDirac2021}.
The NQR spectra measured at 4.2 K under various pressure values are presented in Fig.~\ref{fig:spectra}(a).
The full width at half maximum (FWHM) at ambient pressure is 0.079~MHz. This is about $1/3$ of FWHM in the powder samples \cite{takahashiSWaveSuperconductivityDirac2021},
indicating a much higher quality of the single-crystalline samples. The spectrum at 0 GPa is compared with that of the powder samples in the inset of Fig.~\ref{fig:spectra}(a).
As pressure increases, the peak frequency decreases. This indicates that the electric field gradient around the observed nuclei varied owing to compression.
In addition, the linewidth of the spectrum monotonically increases with applying pressure as shown in Fig.~\ref{fig:spectra}(b).
If the environment around the observed nuclei is homogeneous throughout the compressed sample, the linewidth would not increase.
In general, measurements under pressure are influenced by the inhomogeneity of pressure, which may cause the broadening of the NQR spectra.
When using a piston-cylinder-type cell and Daphne 7373 oil, it was reported that pressure inhomogeneity is small up to 2.2 GPa above which Daphne 7373 oil becomes
solid at room temperature, and that the inhomogeneity is almost constant between 1 and 2.2 GPa \cite{PhysRevB.77.064508,fukazawaEvaluationPressureTransmitting2007}.
Therefore, the observed continuous increase in the FWHM above 1.10 GPa cannot be attributed to the inhomogeneity of pressure. 
Another possibility is plastic deformation of the crystal. We can exclude this for the following reason.
After the compression process, we relieved pressure and measured the NQR spectrum at ambient pressure again.
The shift of the resonance frequency was much smaller than the original FWHM and 
the value of FWHM was almost unchanged compared to the change caused by pressure, indicating the shrinkage is reversible in the pressure region below 2.08 GPa.
We believe that the nonsaturating broadening of the NQR spectrum reflects the unique features of the compression of \casb.
Specifically, a lattice anomaly occurring at higher pressure as seen in Cd$_2$Re$_2$O$_7$ is anticipated \cite{kitagawaVariationSuperconductingSymmetry2020}.
In order to understand the evolution of the structural parameters and electric field gradient against pressure, the detection of all NQR signals is desirable.

\begin{figure}[tbp]
        \includegraphics[width=8.6cm]{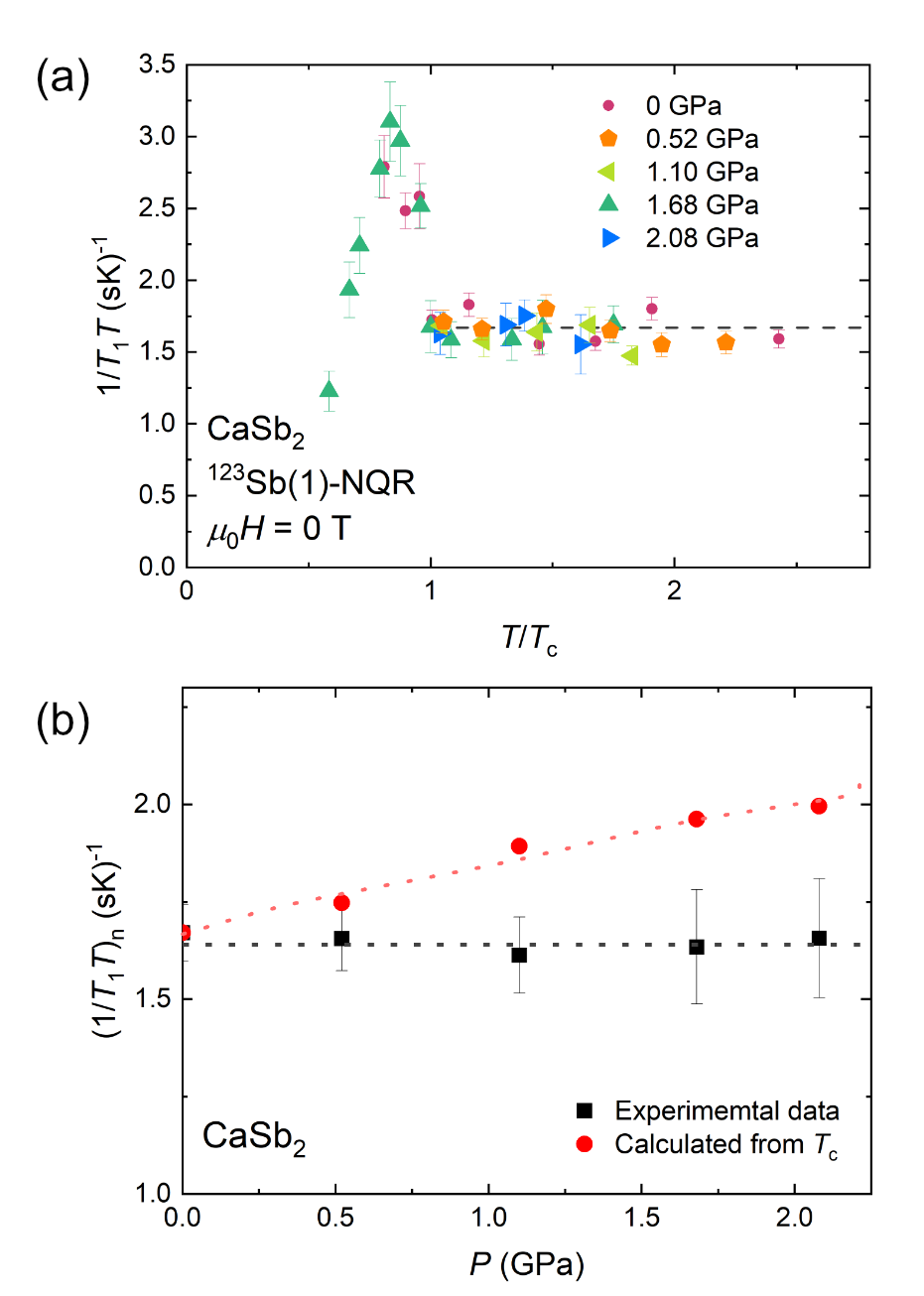}
        \caption{(Color online) 
        (a) $T/T_{\mathrm{c}}$ dependence of $1/T_1T$ under various pressure values.
        (b) Pressure dependence of observed Korringa value (black squares) and calculated ones (red circles).}
        \label{fig:t1}
\end{figure}

Figure \ref{fig:t1}(a) shows the normalized temperature $T/T_{\mathrm{c}}$ dependence of $1/T_1T$ at each pressure.
Above \Tc, ~$1/T_1T$ is almost constant in all pressure values.
This so-called Korringa behavior is a characteristic of Fermi liquid-systems.
Considering $1/T_1T \propto N(E_{\mathrm{F}})^2$, pressure independent $1/T_1T$ indicates that $N(E_{\mathrm{F}})$ is independent of pressure.
This suggests that the increase in \Tc ~cannot be explained by that in $N(E_{\mathrm{F}})$.
We discuss this point later.

\begin{figure}[tbp]
    \includegraphics[width=8.65cm]{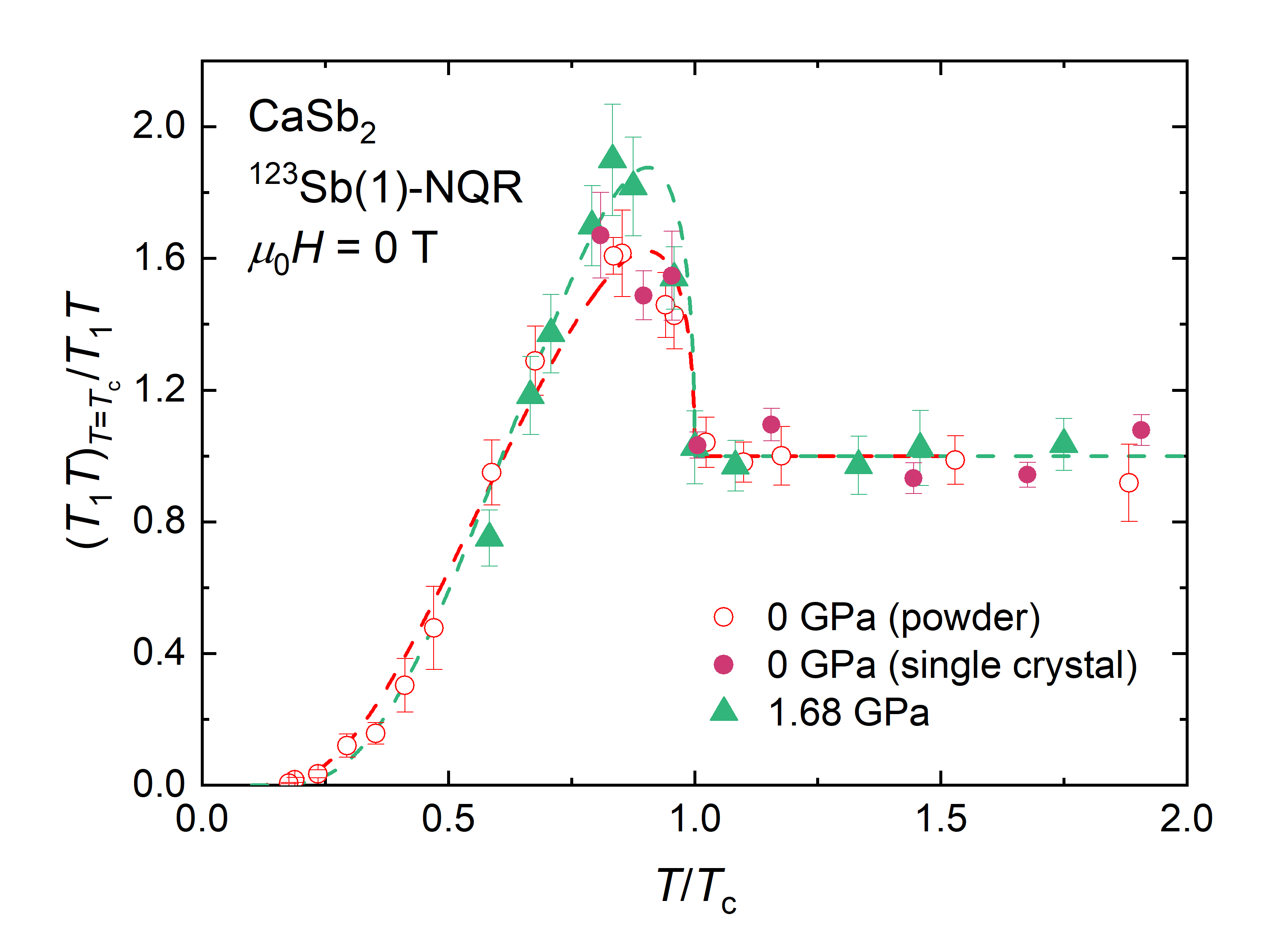}
    \caption{(Color online) 
    $T/T_{\mathrm{c}}$ dependence of normalized $1/T_1T$ in the low-temperature region at ambient pressure and 1.68 GPa. Dashed curves show the fitting results using the BCS model.}
    \label{fig:SCt1}
\end{figure}

We measured $1/T_1T$ below \Tc ~at 0 and 1.68 GPa to compare the SC properties under pressure with those at ambient pressure.
In the SC state, a clear coherence peak was observed just below \Tc ~at 1.68 GPa as shown in Fig.~\ref{fig:t1}(a).
This suggests that BCS superconductivity continues in this pressure region.
We also normalize $1/T_1T$ with the mean value of that above $T_{\mathrm{c}}$ and plot it for 1.68 GPa
along with that at ambient pressure \cite{takahashiSWaveSuperconductivityDirac2021} as shown in Fig.~\ref{fig:SCt1}.
The normalized coherence peak becomes higher under pressure.
We fitted the data using the BCS model \cite{hebelTheoryNuclearSpin1959},
\begin{equation}
    \frac{T_{1\mathrm{n}}}{T_{1\mathrm{s}}}=\frac{2}{k_{\mathrm{B}}T}\int_0^{\infty} dE\ N_{\mathrm{s}}^2(E)\left[1+\frac{|\Delta(T)|^2}{E^2}\right]f(E)\left[1-f(E)\right],
\end{equation}
where $T_{1\mathrm{n}}$ ($T_{1\mathrm{s}}$) is $T_1$ in the normal (SC) state, $N_{\mathrm{s}}(E)$ is the quasiparticle DOS in the SC state, $\Delta(T)$ 
is the $T$-dependent energy gap, and $f(E)$ is the Fermi distribution function.
The effect of the energy broadening in $N_{\mathrm{s}}(E)$ was introduced by the convolution with a rectangular broadening function whose width and height are $2\delta$ and $1/2\delta$,
respectively. $\delta$ corresponds to the distribution of the SC gap.
Although the existence of two SC gaps was discussed from the temperature dependence of the SC upper critical field $H_{\mathrm{c2}}$ \cite{ikedaQuasitwodimensionalFermiSurface2022} and the magnetic penetration depth \cite{PhysRevB.106.214521},
a single gap was sufficient for the fitting of $1/T_1T$ in the SC state. 
(We also performed the fitting with a two-gap model, but the two obtained gaps had nearly the same values.)
The results are shown as dashed curves in Fig.~\ref{fig:SCt1}.
At 1.68 GPa, we obtained $\Delta(0)/k_{\mathrm{B}}T_{\mathrm{c}}=1.72$ and $\delta/\Delta(0)=0.17$.
Compared with the values at ambient pressure [$\Delta(0)/k_{\mathrm{B}}T_{\mathrm{c}}=1.52$ and $\delta/\Delta(0)=0.35$] \cite{takahashiSWaveSuperconductivityDirac2021}, 
the size of the SC gap became larger and the distribution of the gap became smaller.
The distribution of the gap mainly comes from the intrinsic gap anisotropy and the inhomogeneity of the sample.
We note that the data just below \Tc ~measured in the single crystalline samples at ambient pressure were almost the same as those in the powder sample \cite{takahashiSWaveSuperconductivityDirac2021}.
To be more specific, there is little change in the height of the coherence peak, which is a decisive factor for $\delta/\Delta(0)$.
Thus, the reduction of the distribution is not due to the sample quality, but to the suppression of the gap anisotropy under pressure.
The origin of a more isotropic gap by applying pressure is unclear, but it might be related to the pressure dependence of phonons.
We consider that the electron scattering by low-energy phonons might be enhanced because of the pressure-induced structural fluctuations, averaging the electronic states over the Fermi surfaces.
As a result, the effective electron-electron interaction is averaged out and becomes momentum independent, making the SC gap more isotropic.
The enhancement of the electron scattering by phonons can also explain the broadening of the NQR spectrum by the disturbance of the electric field gradient.

Here, we discuss the origin of the enhancement in \Tc ~by applying pressure.
In weak-coupling BCS superconductors, \Tc ~is expressed as
\begin{equation}
    T_{\mathrm{c}}=1.13\frac{\hbar\omega_{\mathrm{D}}}{k_{\mathrm{B}}}\exp\left[-\frac{1}{N(E_{\mathrm{F}})V}\right],\label{BCS}
\end{equation}
where $\omega_{\mathrm{D}}$ is the Debye frequency, $V$ is the effective interaction between electrons mediated by phonons, and $k_{\mathrm{B}}$ is Boltzmann's constant.
Accordingly, $N(E_{\mathrm{F}})$ is one of the parameters governing \Tc.
In conventional metallic systems, $1/T_1$ follows the Korringa law expressed as
\begin{equation}
    \frac{1}{T_1T}=\frac{\pi}{\hbar}A^2N(E_{\mathrm{F}})^2k_{\mathrm{B}},
\end{equation}
where $\hbar$ and $A$ are Dirac's constant and the hyperfine coupling constant, 
respectively.
By measuring the value of $1/T_1T$, we can detect the change in $N(E_{\mathrm{F}})$ and confirm whether or not the increase in \Tc ~is related to $N(E_{\mathrm{F}})$ 
\cite{kotegawaEvidenceHighfrequencyPhonon2002,mukudaRuNQRProbe1998}.
In \casb, we observed a conventional metallic behavior originating from Fermi-surface portions away from the topological line nodes above \Tc, and BCS SC behavior below \Tc
~\cite{takahashiSWaveSuperconductivityDirac2021}. Therefore, we can apply the above method to \casb.
In our experimental results, $1/T_1T$ was unchanged by pressure, indicating that $N(E_{\mathrm{F}})$ is independent of applied pressure.
Note that the increase in $N(E_{\mathrm{F}})$ accompanied by the decrease in the hyperfine coupling, which results in constant $1/T_1T$, is not expected in weak-coupling electron systems.
In most cases, the hyperfine coupling is kept constant or increased by compression in nonmagnetic metals \cite{BENEDEK1958241,PhysRev.178.998, kitagawaSpaceEfficientOpposedAnvil2010}.
Since \casb ~has multiple bands crossing the Fermi energy, if some of the bands have weak hyperfine coupling, those bands may not be seen by NQR. However, the possibility that the pressure-dependent band is essentially invisible is also excluded because $1/T_1T$ at 1.68 GPa shows a clear 
coherence peak just below the enhanced \Tc. Our measurements reliably captured the electronic states changed by pressure.
For a more detailed discussion, we estimated the expected pressure variation of $1/T_1T$ by assuming that the increase in \Tc ~is caused only by that in $N(E_{\mathrm{F}})$.
Here, we used $\hbar\omega_{\mathrm{D}}/k_{\mathrm{B}}=231$ K \cite{PhysRevB.105.184504} and assumed that $\omega_{\mathrm{D}}$ and $V$ are pressure independent. 
The result is shown in Fig.~\ref{fig:t1}(b) with experimental data.
Clearly, there is a discrepancy between calculation and experimental data.
From this result, we conclude that pressure dependence of \Tc ~does not arise from $N(E_{\mathrm{F}})$.
As in Eq.~\eqref{BCS}, \Tc ~is tuned by three parameters $N(E_{\mathrm{F}})$, $\omega_{\mathrm{D}}$, and $V$. 
Therefore, the origin is considered to be the Debye frequency $\omega_{\mathrm{D}}$ or the effective interaction $V$.
This is surprising because in a weak electron-phonon coupling superconductor, the pressure dependence of $V$ and $\omega_{\mathrm{D}}$ is 
usually weaker than that of $N(E_{\mathrm{F}})$ \cite{PhysRev.159.353}.
Regardless of the origin, there should be a change in the lattice properties.
The peak in \Tc ~suggests that the pressure variation of the lattce properties would have an anomaly.
As the superconductivity of \casb ~is likely to be mediated by electron-phonon coupling, this anomaly might cause interesting phenomena 
such as a drastic change of the SC state.
In particular, it is predicted that the SC symmetry other than the $s$ wave in this material is topologically nontrivial \cite{onoMathbbZEnrichedSymmetry2021}.
Another fascinating possibility for the realization of topological superconductivity is the formation of interband pairing. In \casb, the interband coupling could be strong because of the
band degeneracy at the nodal lines. If such coupling becomes stronger by pressure, the interband pairing would be promoted. Notably, \Tc ~is always enhanced when interband pairing is additionally formed \cite{samokhin2023ginzburglandau}.
An experimental investigation that can probe the pressure variation of $\omega_{\mathrm{D}}$ or $V$ such as a specific-heat measurement is desired.

\textit{Conclusions.}
In conclusion, we performed ac magnetic susceptibility and $^{123}$Sb(1) NQR measurements on single-crystal \casb 
~to investigate the origin of the enhancement in \Tc ~by pressure.
We confirmed the enhancement in \Tc ~of the current samples up to 2.08 GPa.
The NQR spectral peak of the single-crystal samples is much sharper than that of the powder samples, ensuring the higher quality of the single crystals.
We observed nonsaturating spectral broadening against pressure that probably reflects the unique compression behavior of \casb, rather than the pressure inhomogeneity.
At all pressure values, $1/T_1$ exhibits Korringa behavior above \Tc.
The constant value of $1/T_1T$ against pressure indicates that $N(E_{\mathrm{F}})$ is independent of pressure.
This result excludes an enhancement of $N(E_{\mathrm{F}})$ from the possible causes of the enhancement in \Tc.
According to the BCS theory, the remaining candidates are $\omega_{\mathrm{D}}$ and $V$, both of which are related to the crystal lattice.
As for the SC properties, the gap becomes larger and more isotropic by applying pressure. 
For further investigations of the pressure evolution of the electronic state, NQR measurements at higher pressures with a different type of cell would be useful.
It is an interesting future issue to clarify the origin of the peak in \Tc ~against pressure in \casb, as well as its SC symmetry in a higher-pressure region.

\textit{Acknowledgments.}
We would like to thank the Research Center for Low Temperature and Materials Sciences, Kyoto University for the stable supply of liquid helium.
This work was supported by Grant-in-Aid for Scientific Research
on Innovative Areas from MEXT, JSPS Core-to-core program (JPJSCCA20170002),
JSPS KAKENHI (No. JP17H06136, No. JP20H00130, No. JP15K21732,
No. JP15H05745, No. JP20KK0061, No. JP19H04696, No. JP19K14657, No. JP20H05158, No. JP21K18600, No. JP22H04933, No. JP22H01168, No. JP23H01124, No. JP22H04473, and No. JP23H04861), ISHIZUE 2023 of Kyoto University Research Development Program, the Murata Science Foundation, the Kyoto University
Foundation, and Toyota Physical and Chemical Research Institute. Research at the University of Maryland was supported by 
the U.S. Department of Energy Award No. DE-SC-0019154 (low temperature characterization), the Gordon and Betty Moore Foundation's EPiQS Initiative through Grant No. GBMF9071 (materials synthesis), 
and the Maryland Quantum Materials Center.
This work was also supported by JST SPRING, Grant No. JPMJSP2110.

\bibliographystyle{apsrev4-2}  
\bibliography{mylist2}

\end{document}